\documentclass[prc,twocolumn,nofootinbib]{revtex4-2}

\usepackage{color}
\usepackage{graphicx}
\usepackage{dcolumn}
\newcolumntype{d}[1]{D{.}{.}{#1}}
\usepackage{bm}
\usepackage{braket}
\usepackage{amsmath}
\usepackage{here}
\usepackage{color}
\usepackage{multirow}

\usepackage{array}
\usepackage{enumerate}
\usepackage{amssymb}
\usepackage{bbm}
\usepackage{dsfont}
\usepackage[scr=rsfs]{mathalpha}
\usepackage{longtable}

\def\be{\begin{equation}} 
\def\ee{\end{equation}}

\usepackage{xcolor}
\graphicspath{{figs/}}
\usepackage[bookmarks=true,colorlinks,
            citecolor=blue,linkcolor=blue,anchorcolor=blue,filecolor=blue,urlcolor=blue,
           ]{hyperref}

\usepackage[normalem]{ulem}

\renewcommand\sout{\bgroup\markoverwith
{\textcolor[rgb]{1,0.75,0.8}{\rule[.5ex]{2pt}{0.8pt}}}\ULon}

\begin{document}

\title{
A microscopic analysis of sub-barrier photo-induced fission in 
$^{236}$U$(\gamma,f)$ based on the non-equilibrium Green function method
}

\author{K. Uzawa}
\affiliation{ 
Nuclear Data Center, Japan Atomic Energy Agency, Tokai 319-1195,  Japan}

\begin{abstract}
Sub-barrier photo-induced fission in $^{236}$U$(\gamma,f)$ is investigated within
the non-equilibrium Green function (NEGF) method.
A model space for the fission process is constructed by superposing Skyrme-Hartree-Fock wave functions
along the fission path allowing the particle-hole excitation.
Then, the transition from the photo-absorption channel to the fission channel is described by
the non-equilibrium Green-function formalism.
The calculated fission cross section 
in the incident gamma-ray energy range
$5 ~ {\rm MeV} \leq E_\gamma \leq 6 ~ {\rm MeV}$ reproduces the overall behavior of the experimental data,
including the suppression below the fission barrier.
An eigenchannel analysis of the wave propagation in the present fission model space
is also performed, and the first eigenchannel is found to dominate the fission probability.
This result supports the Bohr-Wheeler transition-state picture from a microscopic viewpoint.
\end{abstract}

\maketitle

\section{Introduction}
In nuclear fission, a heavy nucleus splits into two smaller fragments.
It is a complex quantum many-body process in which the nucleus undergoes large-amplitude deformation, and the interplay between collective and single-particle motions leads to energy dissipation.
Due to this complexity, understanding the fission process on the basis of microscopic degrees of freedom remains a challenging problem in nuclear physics.

To investigate fission dynamics, the time-dependent density functional theory (TDDFT) has been widely employed \cite{Negele1978, Simenel2014, Goddard2015, Goddard2016, Bulgac2016, Scamps2018}. This approaches naturally capture the large-amplitude deformation after the barrier penetration and incorporate one-body dissipation effects. However, because TDDFT cannot describe quantum tunneling through the fission barrier, they are not applicable to barrier-penetration dynamics or to the calculation of fission probabilities.

As an alternative approach, the time-dependent generator coordinate method (TDGCM) has been developed 
\cite{Berger1991, Goutte2005, Bernard2011, Regnier2016, Zdeb2017, Tao2017, Verriere2020, Zhao2021,Zhao2022, Schunck2023}.
In this framework, the model space is constructed by superposing mean-field wave functions with different deformation parameters, sometimes augmented by additional collective coordinates \cite{Zhao2021} or quasiparticle excitations \cite{Bernard2011, Schunck2023}.
Although the accessible deformations are restricted to those spanned by the preselected basis states, unlike TDDFT, TDGCM is capable of describing tunneling through the fission barrier.

While these time-dependent approaches provide an intuitive picture of fission dynamics,
there remains some arbitrariness in the preparation of the initial states for the time evolution.
In particular, for induced fission reactions, the initial states should be represented as wave packets from the reaction entrance channels.
However, this point has not been treated consistently in the time-dependent framework. This issue becomes important when calculating fission cross sections and analyzing entrance-channel effects.

One possible solution to this problem is provided by the non-equilibrium Green function (NEGF) method \cite{Datta1995, Datta2005}. This method has been widely used in electronic device physics for the microscopic description of electron transport. The application of the NEGF method to nuclear fission was first discussed in Ref.~\cite{Bertsch2020}, where an induced-fission model based on the mixing of the Slater-determinants with different deformation and particle-hole excitation was proposed.
The equivalence of the approach in Ref.~\cite{Bender2020} to the NEGF framework was pointed out later in Ref.~\cite{Alhassid2021}.

In the NEGF method,
transition from the entrance channels to the exit channels via intermediate states
is described by the Green functions
\footnote{In applications of the NEGF method to transistors, the entrance and exit channels correspond to the leads connected to the left and right electron reservoirs, while the intermediate states correspond to the channel material.}.
In modern NEGF calculations, the model space is constructed by superposing Kohn-Sham Slater determinants
\cite{ Damle2001, Brandbyge2002, Ozaki2010, Thoss2018}.
Motivated by this framework, applications of the NEGF method to induced fission reactions combined with nuclear density functional theory (DFT) have also been explored in Refs.~\cite{Bertsch2023, Uzawa2024, Uzawa2024-2, Uzawa2025}.

In contrast to TDDFT and TDGCM, the NEGF method provides a natural framework for calculating fission cross sections, since it is grounded in the compound nucleus reaction theory \cite{Alhassid2021}.
In addition, the NEGF method has the further advantage that single-particle excitations during the fission process can be incorporated naturally by including particle-hole excited Slater determinants in the model space, although this requires a very large number of basis states and considerable numerical effort. 

In this paper, I extend the NEGF-based theoretical framework developed in Ref.~\cite{Uzawa2025}, which was applied to the $^{235}$U$(n,f)$ reaction, and analyze the $^{236}$U$(\gamma,f)$ reaction. In the $^{235}$U$(n,f)$ reaction, the excitation energy of the $^{236}$U compound nucleus is higher than the fission barrier height, and quantum tunneling is therefore not important. In contrast, in the $^{236}$U$(\gamma,f)$ reaction, the sub-barrier energy region can be accessed by selecting the incident photon energy. I thus examine the applicability of the NEGF approach in the sub-barrier region and analyze the fission mechanism in this low-energy regime. 

This paper is organized as follows. In Sec.~II, I present the theoretical formulation of the NEGF method for photo-induced fission reactions. In Sec.~III, I present the results for the fission cross section of the $^{236}$U$(\gamma,f)$ reaction and compare them with experimental data. I also present an analysis of the reaction process based on the eigenchannel decomposition method. Finally, in Sec.~IV, I summarize the paper.

\section{Theoretical formulation}

In this section, I summarize the theoretical framework and numerical setup of the NEGF calculation for the photo-induced fission reaction $^{236}$U$(\gamma,f)$. The choice of the Skyrme energy density functional, the residual interactions, and the model space is the same as in Ref.~\cite{Uzawa2025}, 
in which the $^{235}$U$(n,f)$ reaction was studied. 
However,
the entrance channels differ between the $^{235}$U$(n,f)$ and $^{236}$U$(\gamma,f)$ reactions, 
and the treatment of the photo-absorption channel is detailed below.

\subsection{Fission path and GCM basis functions}\label{model}

To describe nuclear deformation and single-particle excitations in fission process,
I adopt the Skyrme-Hartree-Fock wave functions $|Q,E_\mu\rangle$, labeled by $Q$ and $E_\mu$, as many-body basis states:
\begin{equation}
   |\Psi\rangle=\int_{Q_{\min}}^{Q_{\max}} dQ \sum_\mu \, f(Q,E_\mu)\, |Q,E_\mu\rangle .
   \label{GCM}
\end{equation}
Here, $Q$ denotes the deformation parameter,
$\mu$ labels the particle-hole excited configurations at $Q$, and
$E_\mu$ is the particle-hole excitation energy measured from the reference state at each $Q$, namely $|Q,E_\mu=0\rangle$. 
$f(Q,E_\mu)$ is the weight function and determined by solving the Hill-Wheeler equation \cite{ring}.

In the following calculations, 
I consider the two-dimensional potential energy surface  of $^{236}$U in the $Q_{20}$-$Q_{30}$ space. 
To obtain the potential energy surface,
I solve the constrained Hartree-Fock equation using the SkyAx code \cite{Reinhard2021}
with the UNEDF1 functional \cite{Kortelainen2012}. 
The resulting potential energy surface is the same as that shown in Fig.~1 of Ref.~\cite{Uzawa2025}.
In that figure,
the fission path is shown by the green line,  
and $Q$ in Eq.~(\ref{GCM}) represents the deformation coordinate along this fission path. 
Pairing correlations are not included at this stage, 
though being incorporated later as residual interactions as Eq.(\ref{pair}).

The discretization of $Q$ is determined by the condition
\begin{equation}
\langle Q_i,E_\mu=0\,|\,Q_{i+1},E_\mu=0\rangle = 0.52,
\end{equation}
to avoid numerical instabilities associated with the overcompleteness of the non-orthogonal basis states.
The selected 19 reference states are indicated by the white circles in the same figure.
For $Q_{\min}$ and $Q_{\max}$ in Eq.(\ref{GCM}), I adopt the same values as in Ref.~\cite{Uzawa2025}:
$(Q_{20},Q_{30})=(14\,\mathrm{b},0)$ for $Q_{\min}$ and  
$(Q_{20},Q_{30})=(84\,\mathrm{b},19\,\mathrm{b}^{3/2})$ for $Q_{\max}$, respectively.

\begin{figure}
\centering
\includegraphics[width=8.6cm]{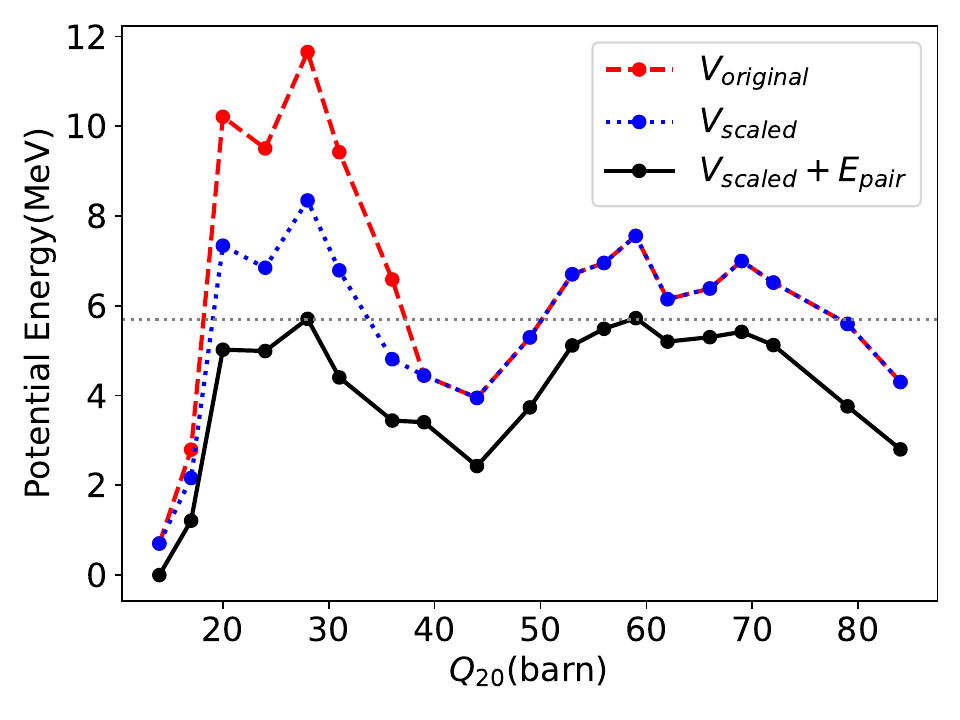}
\caption{
The fission barrier along the fission path shown in Fig.1 in Ref.~\cite{Uzawa2025}. 
The red-dashed line
shows the original barrier, while the blue-dotted line is scaled by
a factor of $f = 0.70$. 
The black-solid line shows the sum of scaled barrier and the paring energy.
The points denote the reference states used for
the GCM calculations.
}
\label{barrier}
\end{figure}

The deformation energy of the selected reference states along the fission path is shown by the red-dashed line in Fig.~\ref{barrier}.
Because pairing correlations and triaxial deformation are not taken into account, the fission barrier height overestimates the experimental value, about 5.7 MeV\cite{VANDENBOSCH1973, Lynn1980, Leal1999}. 
I therefore renormalize the first fission barrier, as indicated by the blue-dotted line in Fig.~\ref{barrier}. 
After this renormalization, the sum of the barrier height and $E_{\rm pair}$ gives an energy difference of 5.7 MeV between $Q_{20}=14$ b and $Q_{20}=28$ b, which is consistent with the experimental fission barrier height.
Note that the empirical fission barrier height contains some uncertainty, depending on the functional form adopted for the fit to the fission cross section and on the fitting procedure.
In the present work, I adopt a value of 5.7 MeV, as in Ref.~\cite{Uzawa2025}.
The determination of $E_{\rm pair}$ requires configuration mixing of the Hartree-Fock basis functions, and will therefore be discussed later in Sec.~\ref{HandN}.

In addition to the discretization of the fission path, the summation over $\mu$ in Eq.~(\ref{GCM}) is truncated according to the energy-cutoff condition $E_\mu \leq E_{\max}(Q)$, defined as \cite{Uzawa2025}
\begin{equation}
E_{\max}(Q)=
\begin{cases}
E_{\max}-V_{\mathrm{scaled}}(Q), & \text{for finite seniority},\\[4pt]
E_{\max}, & \text{for zero seniority}.
\end{cases}
\end{equation}
The cutoff parameter $E_{\max}$ must be taken sufficiently large.
In Ref.~\cite{Uzawa2025}, it was shown that $E_{\max} \geq 7.0$ MeV ensures convergence of the fission cross section for $^{235}$U$(n,f)$ at $E_n=10$ keV. 
In the present study, I consider incident $\gamma$-ray energies in the range $5\,\mathrm{MeV} \leq E_\gamma \leq 6\,\mathrm{MeV}$, for which the excitation energy of the compound nucleus is lower than that in the neutron-induced case. 
I therefore adopt $E_{\max}=7.0$ MeV in the present calculation.

From the generated particle-hole excited configurations,
I retain only those with $K=0$, where $K$ is the projection of the angular momentum onto the symmetry axis,
since $K$ is a good quantum number in the present axially symmetric calculation.
As a result, the total number of selected configurations is $N_{\mathrm{dim}}=145124$, and these states are used as the basis for constructing the overlap and Hamiltonian matrices in the next subsection.

\subsection{Overlap and Hamiltonian matrices}\label{HandN}
Based on the many-body basis in Eq.(\ref{GCM}),
the overlap and Hamiltonian matrix elements are calculated as
\begin{equation}
    N_{ij}=\langle Q,E_\mu \mid Q',E_{\mu'}\rangle,
\end{equation}
\begin{equation}
    H_{ij}=\langle Q,E_\mu \mid H \mid Q',E_{\mu'}\rangle.
\end{equation}
Here the index $i$  represents both $Q$ and $E_\mu$, and same for $j$.

The diagonal elements of $N_{ij}$ are equal to unity, while the off-diagonal elements are evaluated approximately following Refs.~\cite{Bertsch2023, Uzawa2024, Uzawa2025}. Assuming that the change in the single-particle wave functions between neighboring deformation points $Q_i$ and $Q_{i+1}$ is small, I approximate
\begin{equation}
\langle Q_i,E_\mu \mid Q_{i+1},E_{\mu'}\rangle = \langle Q_i \mid Q_{i+1}\rangle
\label{OLP_approx}
\end{equation}
for diabatically connected configurations, and set it to zero otherwise. By construction of the discretized basis, $\langle Q_i \mid Q_{i+1}\rangle = 0.52$ for all $i$. 
Within the Gaussian overlap approximation (GOA) \cite{ring}, the overlap decreases in a Gaussian manner as the distance between basis states increases. I therefore neglect couplings between next-nearest neighboring deformation points.

For the Hamiltonian matrix, the diagonal elements,
$\langle Q,E_\mu|  H | Q,E_\mu\rangle$,
are given by the sum of the deformation potential energy $V(Q)$ and the particle-hole excitation energy $E_\mu$.

For the off-diagonal matrix elements, I take into account three types of residual interactions that play important roles in fission dynamics.
The first is the monopole-type pairing interaction \cite{Barranco1990},
\begin{equation}
    H_{\mathrm{pair}}
    =
    -G\sum_{i\neq j}
    a_i^\dagger a_{\bar{i}}^\dagger a_{\bar{j}} a_j.
    \label{pair}
\end{equation}
The strength parameter $G$ is determined so as to reproduce the first excited $0^+$ state, as in Ref.~\cite{Uzawa2025},
and I use $G=0.1465$ MeV.

In addition, I introduce a random residual interaction of the form
\begin{equation}
    H_{\rm ran}=v\sum' r_{ijkl}\, a_i^\dagger a_j^\dagger a_l a_k,
    \label{ran}
\end{equation}
where $r_{ijkl}$ denotes a random number sampled from the standard normal distribution, and the summation is restricted to configurations that conserve the $K$ quantum number.
The interaction strength was estimated in Ref.~\cite{Uzawa2023}, yielding $v=0.025$ MeV and $v=0.02$ MeV for the proton-neutron and proton-proton (neutron-neutron) channels, respectively. This random residual interaction, particularly in the proton-neutron channel, was shown to play an important role in dissipative shape evolution in Ref.~\cite{Bush1992} .

Furthermore, I include the diabatic interaction introduced in Ref.~\cite{Hagino2022}:
\begin{eqnarray}
\frac{\langle Q,E_\mu|v_{\rm db}|Q',E_{\mu'}\rangle}{\langle Q,E_\mu|Q',E_{\mu'}\rangle}
&=&\frac{E(Q,E_\mu)+E(Q',E_{\mu'})}{2} \nonumber \\
&&+ h_2 \ln \!\left(\langle Q,E_\mu|Q',E_{\mu'}\rangle\right).
\label{eq:H_db}
\end{eqnarray}
This term connects diabatically connected configurations with $\langle Q,E_{\mu}|Q',E_{\mu'}\rangle \neq 0$ and originates from the second-order term in the GOA.
The value of $h_2$ was estimated to be $1.5$ MeV for $^{236}$U in Ref.~\cite{Bertsch2023}, and the same value is adopted in the present work.

$E_{\rm pair}(Q)$ in Fig.~\ref{barrier} is defined as the lowest eigenvalue of the matrix
$\langle Q,E_\mu|H_{\rm pair}|Q,E_{\mu'}\rangle$
constructed within each fixed-$Q$ subspace.
At $Q_{\rm min}$, $E_{\rm pair}=-0.703$ MeV,
and all excitation energies in the following are measured relative to this reference value.

\subsection{Decay width matrices}
\subsubsection{Photon transmission coefficients}\label{T_photo}

The goal in this paper is to describe the photo-induced fission within the NEGF framework. To reduce the ambiguity in the treatment of the photon-nucleus interaction, I determine the photo-absorption and photo-emission widths, $\Gamma_{\gamma_{\rm in}}$ and $\Gamma_{\gamma_{\rm out}}$, from empirical values \cite{ripl, iwamoto2016}.
In addition, for simplicity, I consider only the $E1$ photon channel and neglect the $M1$, $E2$, and higher-multipole channels, whose contributions are much smaller than that of the $E1$ channel.

Using the $E1$ gamma strength function, $f_{E1}$,
the transmission coefficient for the $E1$ photon
is expressed as \cite{ripl, iwamoto2016}
\begin{equation}
    T_{\gamma_{\rm in}}(E_\gamma)=2\pi E_\gamma^3f_{E1}(E_\gamma).
    \label{T_g_in}
\end{equation}
Under the Brink-Axel hypothesis\cite{Axel1962},
the transmission coefficient for the $E1$ photon emission from an excited nucleus with excitation energy $E$ is
\begin{equation}
    T_{\gamma_{\rm out}}(E)=\sum_{J\pi}\int^E_0 dE_\gamma \rho(E-E_\gamma,J,\pi)T_{\gamma_{\rm in}}(E_\gamma),
    \label{T_g_out}
\end{equation}
where $\rho(E,J,\pi)$ denotes the level density with spin $J$ and parity $\pi$, and the sum runs over $J$ and $\pi$ values that satisfy the selection rules for $E1$ photon emission.

For practical calculations, the functional forms of $f_{E1}(E)$ and $\rho(E,J,\pi)$ need to be specified.
The recommended parameters are summarized in RIPL-3 \cite{ripl} for the Lorentzian form of $f_{E1}(E)$ :
\begin{equation}
    f_{E1}(E)=\sum_{i=1}^{2}\sigma_i \Gamma_i
    \frac{E\Gamma_i}{(E^2-E_i^2)^2+(E\Gamma_i)^2}.
    \label{GSF}
\end{equation}
The values of the GDR peak energies $E_i$, widths $\Gamma_i$, and strengths $\sigma_i$ are summarized in Table~\ref{table:fE1}.

\begin{table}[htbp]
\centering
\caption{Empirical GDR parameters for $^{236}$U.}
\begin{tabular}{cccc}
\hline
$i$ & $E_i$ (MeV) & $\sigma_i$ (mb) & $\Gamma_i$ (MeV) \\
\hline
1 & 10.93 & 278.4 & 2.58 \\
2 & 13.80 & 410.3 & 4.78 \\
\hline
\end{tabular}
\label{table:fE1}
\end{table}

For the level density $\rho(E,J,\pi)$, I adopt the Gilbert--Cameron formula, in which the constant-temperature and Fermi-gas models are smoothly connected at a matching energy \cite{Gilbert1965}. This model has the several fitting parameters and the details of the parameter determination are summarized in Appendix~\ref{app.,LD}.

\begin{figure}
\centering
\includegraphics[width=8.6cm]{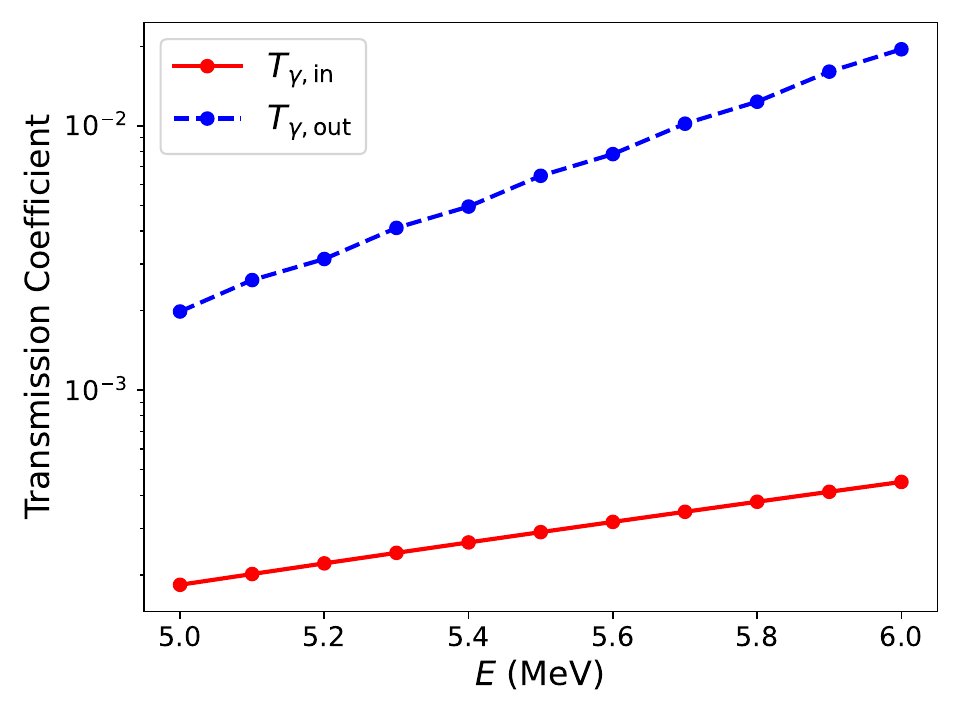}
\caption{
The photo-absorption and photo-emission transmission coefficients defined in Eqs.~(\ref{T_g_in}) and (\ref{T_g_out}). The $E1$ gamma strength function and the level density defined in sec.\ref{T_photo} are used.
}
\label{TC}
\end{figure}

The resulting transmission coefficients for photon absorption and photon emission are shown in Fig.~\ref{TC}.

\subsubsection{Gamma width}

From the photon transmission coefficients, 
the average photo-absorption and photo-emission widths  are obtained as
\begin{equation}
    \langle \Gamma_a \rangle(E) = \frac{T_a(E)}{2\pi \rho(E)}, ~(a=\gamma_{\rm in}, \gamma_{\rm out})
    \label{CN_width}
\end{equation}
under the statistical model assumption \cite{Feshbach1958}.
Using these quantities, I define the decay-width matrices by
\begin{equation}
    (\Gamma_{\gamma_{\rm in}})_{ij}
    =
    \langle \Gamma_{\gamma_{\rm in}} \rangle
    \sum_{k\in \gamma_{\rm in}} N^{1/2}_{k,i} N^{1/2}_{k,j},
    \label{Gg_in}
\end{equation}
and
\begin{equation}
    (\Gamma_{\gamma_{\rm out}})_{ij}
    =
    \langle \Gamma_{\gamma_{\rm out}} \rangle
    \sum_{k\in \gamma_{\rm out}} N^{1/2}_{k,i} N^{1/2}_{k,j},
    \label{Gg_out}
\end{equation}
respectively. The square-root of the overlap matrix is introduced to make the decay width matrices diagonal after the basis orthogonalization \cite{Uzawa2023}:
\begin{equation}
    H\rightarrow N^{-1/2}HN^{-1/2}\equiv\tilde{H}. 
    \label{Lowdin}
\end{equation}

The index $k$ in Eqs.~(\ref{Gg_in}) and (\ref{Gg_out}) runs over the states coupled to the photo-absorption and photo-emission channels.
The number of photo-emission channels is estimated to be of the order of several tens 
based on the Porter-Thomas statistical analysis \cite{Leal1999} .
In the present calculation, I select 100 particle-hole configurations at $Q_{\min}$ whose particle-hole excitation energies are close to the incident photon energy $E_\gamma$, and couple them to the $\gamma_{\rm in}$ and $\gamma_{\rm out}$ channels. These configurations are included in Eqs.~(\ref{Gg_in}) and (\ref{Gg_out}). 
Note that, by normalizing the magnitude of $\langle \Gamma_a \rangle$ by the number of included channels (100 in the present case), the mean reaction probability remains unchanged.

\subsubsection{Fission width}
As in Ref.~\cite{Uzawa2025},
the fission width matrix is defined as
\begin{equation}
(\Gamma_{\rm fis})_{ij}
= \langle \Gamma_{\rm fis} \rangle 
\sum_{k \in {\rm fis}} N^{1/2}_{k,i} N^{1/2}_{k,j},
\label{Gf}
\end{equation}
where the summation $\sum_{k \in {\rm fis}}$ is taken over all configurations at $Q_{\rm max}$.
In contrast to $\langle \Gamma_{\gamma_{\rm in}} \rangle$ and $\langle \Gamma_{\gamma_{\rm out}} \rangle$, $\langle \Gamma_{\rm fis} \rangle$ cannot be estimated from empirical transmission coefficients, because empirical fission transmission coefficients are defined for configurations before the fission barrier, not for those beyond the barrier.
On the other hand, the fission probability is known to be insensitive to the magnitude of $\langle \Gamma_{\rm fis} \rangle$ \cite{Bertsch2023, Uzawa2023, Uzawa2025}. Following Ref.~\cite{Uzawa2025}, I therefore adopt $\langle \Gamma_{\rm fis} \rangle = 100$ keV, which was shown to provide converged results for the same model space as used in the present study.

\subsubsection{Normalization for the decay width}

\begin{figure}
\centering
\includegraphics[width=8.6cm]{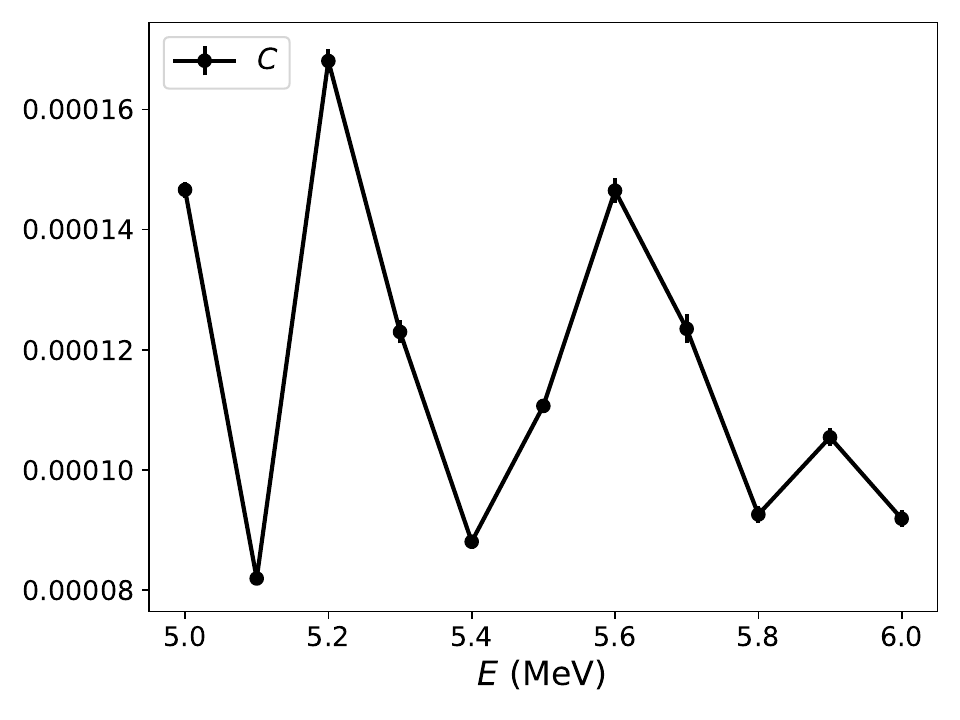}
\caption{
Normalization factor $C$ in Eq. (\ref{norm}) for the photo-absorption or the photo-emission channel as a function of $E$.
The values are averaged over 960 GCM eigenstates.
}
\label{normal}
\end{figure}

Using the overlap matrix $N$, the Hamiltonian matrix $H$, and the decay width matrix $\Gamma\equiv\Gamma_{\gamma_{\rm in}}+\Gamma_{\gamma_{\rm out}}+\Gamma_{\rm fis}$,
the non-Hermitian Hill-Wheeler equation is given by,
\begin{equation}
    \left(H-\frac{i}{2}\Gamma\right)f_\lambda=\tilde{E}_\lambda Nf_\lambda.
    \label{HWeq}
\end{equation}
Here 
$\lambda$ is the label of the eigenstates
and the eigenvalue $\tilde{E}_\lambda$ is the complex number.
The imaginary part of $\tilde{E}_\lambda$ is expected to 
satisfy
\begin{equation}
    -2\,\mathrm{Im}(\tilde{E}_\lambda)=\langle \Gamma_{\gamma_{\rm in}} \rangle +\langle \Gamma_{\gamma_{\rm out}} \rangle +\langle \Gamma_{\rm fis} \rangle ,
\end{equation}
on average.
In the first order pertubation,
the imaginary part for the eigenstates $\lambda$ is written as
\begin{equation}
    \langle \lambda|\Gamma_{a}|\lambda\rangle
    = \langle \Gamma_{a} \rangle
    \sum_{k\in a} |g_\lambda(k)|^2.
    \label{G_pert}
\end{equation}
Here $g_\lambda$ is the collective wave function defined by 
\begin{equation}
    Hf_\lambda=E_\lambda Nf_\lambda,
\end{equation}
and
\begin{equation}
    g_\lambda = N^{1/2} f_\lambda.
\end{equation}
The latter factor in eq. (\ref{G_pert}),
\begin{equation}
    C \equiv \sum_{k\in a} |g_\lambda(k)|^2,
    \label{norm}
\end{equation}
corresponds to the sum of the spectroscopic factors for channel $a$.
On the other hand, 
in the empirical fitting of the Gamma strength function in eq. (\ref{GSF}),
the contribution of the spectroscopic factor is already included.
Therefore, for consistency,
I have to renormalize the decay width as
\begin{equation}
    \langle \Gamma_{a} \rangle \rightarrow \langle \Gamma_{a} \rangle/C.
\end{equation}
Figure~\ref{normal} shows the values of $C$ for the $\gamma_{\rm in}$ and $\gamma_{\rm out}$ channels, where the ensemble average is taken over 960 eigensolutions of the Hermitian Hill--Wheeler equation with eigenenergies within 10 keV of $E$. The value of $C$ exhibits an overall decreasing trend, together with fluctuations reflecting the structure of the selected basis functions.

On the other hand, no additional normalization is introduced for the fission width, since the adopted value, 
$\langle \Gamma_{\rm fis}\rangle=100$ keV, was determined in Ref.~\cite{Uzawa2025} with the normalization effect already taken into account.

\subsection{Fission cross section}
Within the NEGF formalism, the retarded Green function is given by \cite{Bertsch2023}
\begin{equation}
    G(E)=\left(NE-H+\frac{i}{2}\Gamma\right)^{-1}.
    \label{inv}
\end{equation}
Using the Green-function matrix and the partial decay-width matrices, the transmission coefficient from channel $a$ to channel $b$ leads
\begin{equation}
T_{ab}(E) = \mathrm{Tr}\left[\Gamma_a G(E)\Gamma_b G^\dagger(E)\right].
\label{Datta}
\end{equation}
I focus on the photo-induced fission transmission coefficient, $T_{\gamma_{\rm in},\mathrm{fis}}(E)$.
The photo-induced fission cross section is then given by the product of the kinematical factor and the transmission coefficient \cite{Beard2012},
\begin{equation}
\sigma_{\gamma_{\rm in},{\rm fis}}(E_\gamma)=\frac{3(\pi\hbar c)^2}{2\pi E^2_\gamma}T_{\gamma_{\rm in},\mathrm{fis}}(E_\gamma).
\label{sigma_T}
\end{equation}

In the following discussion, I focus on the photon energy region $5 \leq E_\gamma \leq 6$ MeV. As discussed in Sec.~\ref{model}, I adopt a fission barrier height of 5.7 MeV, and this photon-energy range covers both the sub-barrier and above-barrier regions.
In this photon energy region, the neutron-emission channel is closed and the resonances are well separated, namely, $\Gamma < 1/\rho$. On the other hand, the experimentally observed photo-induced fission cross section is effectively energy-averaged due to the finite energy resolution and the unfolding procedure associated with the incident photon-energy distribution \cite{Caldwell1980, Yester1986, Soldatov1995}. Therefore, on the theoretical side, an energy average of the cross section is required.

To evaluate the matrix elements of the Green function appearing in the trace formula of Eq.~(\ref{Datta}), one must solve the large-dimensional linear equation (\ref{inv}) with $N_{\rm dim}=145124$. For the energy average of the transmission coefficients or the cross sections, the Green function needs to be computed on a fine energy mesh comparable to or smaller than the resonance width, $\Gamma \sim O(10^{-8})$, over an energy interval characterized by the mean level spacing, $1/\rho \sim O(10^{-4})$. Repeatedly solving such a large-dimensional linear equation on so fine an energy mesh is computationally prohibitive.

To overcome this difficulty, we have developed an alternative method
for evaluating the Green function matrix based on the
shift-invert Lanczos or Arnoldi methods \cite{Uzawa2024-2}.
This approach uses the spectral representation of the Green
function and utilizes the eigensolutions of the Hill-Wheeler Eq. in (\ref{HWeq}).
Once the eigensolutions are obtained, the Green function can be evaluated for different values of $E$ without repeatedly solving the large-dimensional linear equation.
The main difference between the Lanczos and Arnoldi methods lies in their treatment of the non-Hermitian part of the Hamiltonian: in the Lanczos method, the Hermitian eigenvalue problem is solved and the non-Hermitian part is treated pertubatively, whereas the Arnoldi method solve the non-Hermitian eigenvalue problem explicitly. 
Although the approximate treatment based on the Lanczos method was shown in Ref.~\cite{Uzawa2024-2} to provide good accuracy, I adopt the more accurate shift-invert Arnoldi method in the present work.

\subsection{Eigenchannel Analysis}
To extract an essential degrees of freedom from wave propagation between the input and output channels, the eigenchannel decomposition of the Green function can be applied.
This method is widely used in nano-device physics \cite{Brandbyge2002, Paulsson2007}.
In addition, 
although the Green function does not explicitly appear,
an equivalent analysis has been applied to heavy-ion fusion reactions
to investigate coupled-channel effects~\cite{Hagino2012}.

In this method, the trace formula is rewritten as
\begin{equation}
\begin{aligned}
T_{ab}
&= \mathrm{Tr}\!\left[\Gamma_a G(E)\Gamma_b G^\dagger(E)\right] \\
&= \mathrm{Tr}\!\left[\Gamma_a^{1/2} G(E)\Gamma_b G^\dagger(E)\Gamma_a^{1/2}\right] \\
&= \mathrm{Tr}\!\left[t t^\dagger\right] \ \ \\
&= \sum_n |t_n|^2 ,
\end{aligned}
\label{EIGEN}
\end{equation}
where the rectangular matrix $t$ is defeined as $t = \Gamma_a^{1/2} G(E)\Gamma_b^{1/2}$.
In the last line of Eq.~(\ref{EIGEN}), the singular value decomposition of $t$ is employed:
\begin{equation}
    t = U\,\mathrm{diag}(t_1,t_2,\ldots)\,V^\dagger,
    \label{SVD}
\end{equation}
where $t_n$ denotes the singular values of $t$. The quantity $T_n \equiv |t_n|^2$ represents the transmission coefficient for the $n$th eigenchannel. The left and right singular vectors, represented by $U$ and $V$, describe how each eigenchannel is distributed over the entrance channel $a$ and the exit channel $b$, respectively. In the following, I analyze the properties of both the singular values and the singular vectors in the photo-induced fission process.

\section{Results}
\subsection{photo-induced fission cross sections}

\begin{figure}
\centering
\includegraphics[width=8.6cm]{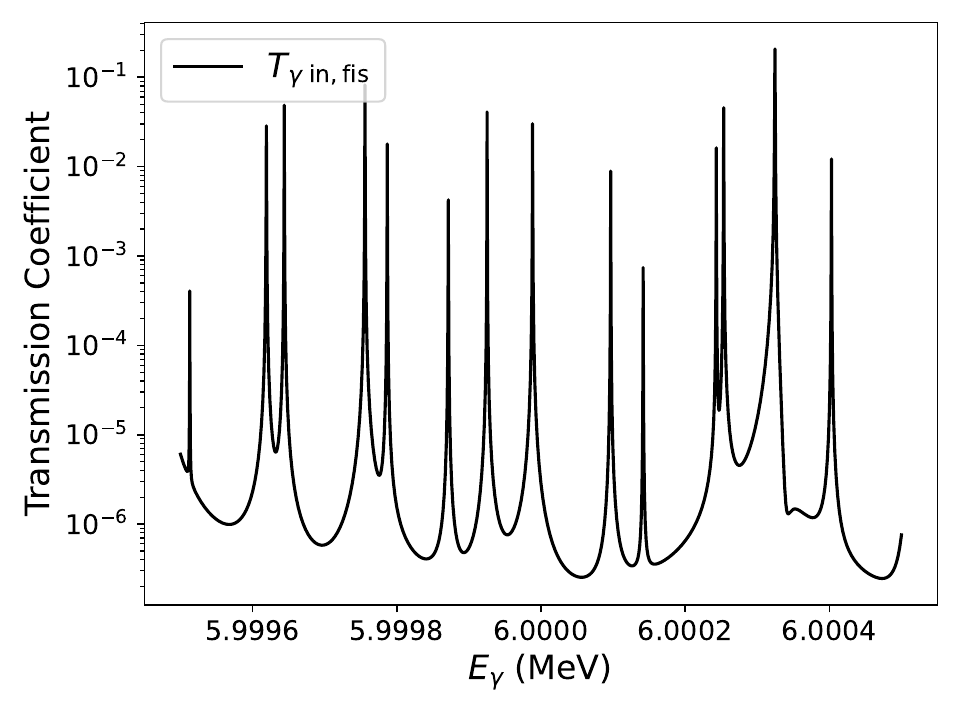}
\caption{
The photo-induced fission transmission coefficient obtained with a specific random seed. Neither the energy average nor the ensemble average is performed.
}
\label{T_resolve}
\end{figure}

In this subsection, I analyze the photo-induced fission cross section in the energy range \(5\,\mathrm{MeV} \leq E_\gamma \leq 6\,\mathrm{MeV}\). First, I calculate the matrix elements of the Green function $G(E_\gamma)$ in Eq.~(\ref{Datta}) and the photo-induced fission transmission coefficient $T_{\gamma_{\rm in},\mathrm{fis}}(E_\gamma)$. Figure~\ref{T_resolve} shows $T_{\gamma_{\rm in},\mathrm{fis}}$ around $E_\gamma = 6$ MeV for a specific random seed. The poles of the Green function correspond to the resonances shown in Fig.~\ref{T_resolve}, and they are resolved in this energy region.

Next, 
in order to compare with the experimental data,
the energy-averaged transmission coeffients
\begin{equation}
\langle T_{\gamma_{\rm in},\mathrm{fis}} \rangle (E_\gamma)
=
\frac{1}{\Delta E}
\int_{E_\gamma-\Delta E/2}^{E_\gamma+\Delta E/2}
T_{\gamma_{\rm in},\mathrm{fis}}(E')\, dE',
\end{equation}
is calculated,
where $\Delta E = 1$ keV and $dE'=10^{-8}$ MeV are used for the energy averaging width and the discretization mesh, respectively.
In addition, due to the usage of the random residual interaction in Eq.(\ref{ran}),
an ensemble average of the enegy averaged transmission coefficents over 96 random seeds is also performed. 

\begin{figure}
\centering
\includegraphics[width=8.6cm]{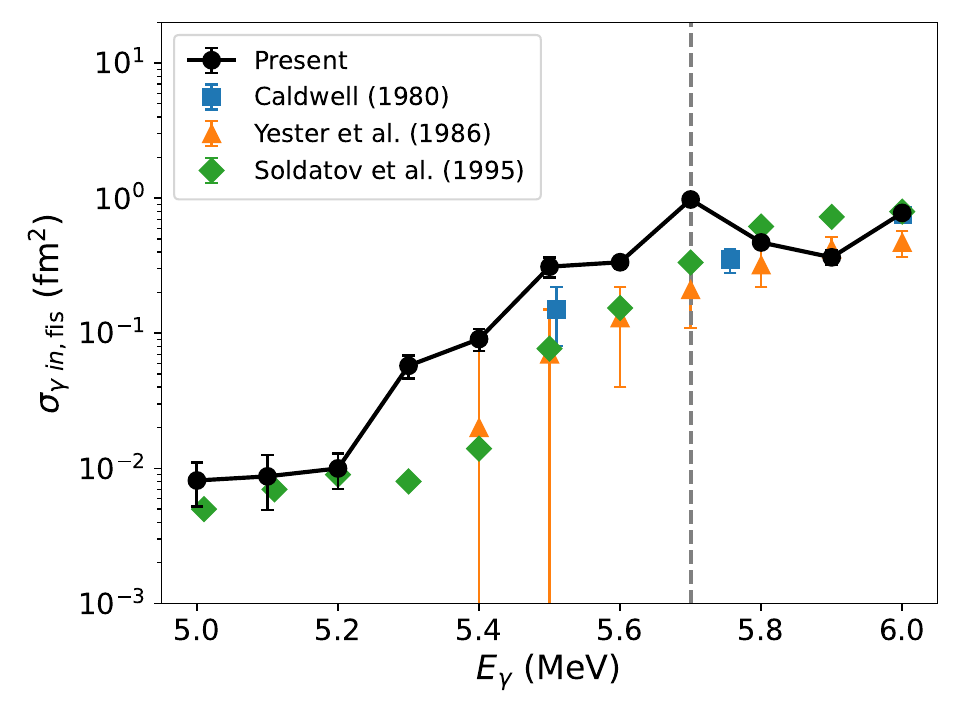}
\caption{
Photo-induced fission cross section calculated with the NEGF formalism (black solid line).
The experimental data are additionally plotted \cite{Caldwell1980, Yester1986, Soldatov1995}.
The vertical line indicates the fission-barrier height of \(5.7\) MeV.
}
\label{sigma}
\end{figure}

The averaged fission cross section,
\begin{equation}
\sigma_{\gamma_{\rm in},\mathrm{fis}}(E_\gamma)
=
\frac{3(\pi\hbar c)^2}{2\pi E_\gamma^2}
\langle T_{\gamma_{\rm in},\mathrm{fis}}\rangle(E_\gamma)
\end{equation}
is shown in Fig.~\ref{sigma}, together with the experimental data. Above the barrier, $E_\gamma \geq 5.7\,\mathrm{MeV}$, the NEGF results agree with the experimental values within a factor of 5. In the sub-barrier region, the calculation reproduces the overall decreasing trend, although the cross section is somewhat overestimated in the range $5.3\,\mathrm{MeV} \leq E_\gamma \leq 5.7\,\mathrm{MeV}$. In contrast, the agreement is very good in the deep sub-barrier region, $5.0\,\mathrm{MeV} \leq E_\gamma \leq 5.2\,\mathrm{MeV}$, where the calculated cross section agrees with the experimental data within a factor of 2.

\begin{figure}
\centering
\includegraphics[width=8.6cm]{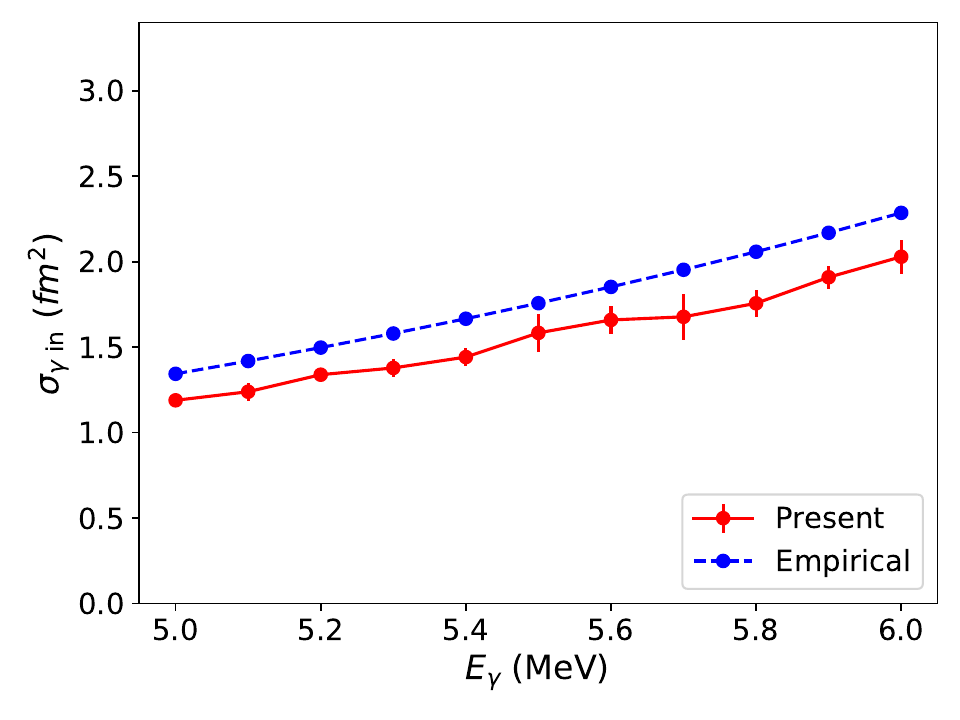}
\caption{
The red solid line shows the photo-absorption cross section obtained from the NEGF calculation, as defined in Eq. (\ref{tot_photo}).
The blue dotted line shows the empirical photo-absorption cross section, given by the product of the transmission coefficient in Eq. 
(\ref{T_g_in}) and the kinematic factor in 
Eq. (\ref{sigma_T}).
}
\label{sigma_abs}
\end{figure}

Note that, in this model, the photo-absorption cross section
\begin{equation}
\sigma_{\gamma_{\rm in}}
\equiv
\sigma_{\gamma_{\rm in},\gamma_{\rm in}}
+\sigma_{\gamma_{\rm in},\gamma_{\rm out}}
+\sigma_{\gamma_{\rm in}, \mathrm{fis}}
\label{tot_photo}
\end{equation}
is adjusted to reproduce a realistic magnitude, since
$T_{\gamma_{\rm in}}$
is determined from the empirical value.
Figure~\ref{sigma_abs} compares the photo-absorption cross section
$\sigma_{\gamma_\mathrm{in}}$ calculated with the NEGF method based on Eq.~(\ref{tot_photo})
with that obtained from the empirical
$T_{\gamma_\mathrm{in}}$ in Eq.~(\ref{T_g_in}).
The two results agree within a factor of 1.2,
indicating that the photon absorption cross section is reproduced reasonably well in the present framework. 
The small deviation in Fig. \ref{sigma_abs} is likely due to the perturbative treatment of the decay width $\Gamma$ for the calculation of $C$ in Eq.(\ref{norm}).
Although the photo-absorption probability is adjusted so as to reproduce the empirical value, the fission probability after photo-absorption is calculated microscopically through the competition with the photo-emission channels. Figure~\ref{sigma} shows that the resulting fission probability is in reasonable agreement with the experimental data.

\subsection{Eigenchannel analysis}

\begin{figure}
\centering
\includegraphics[width=8.6cm]{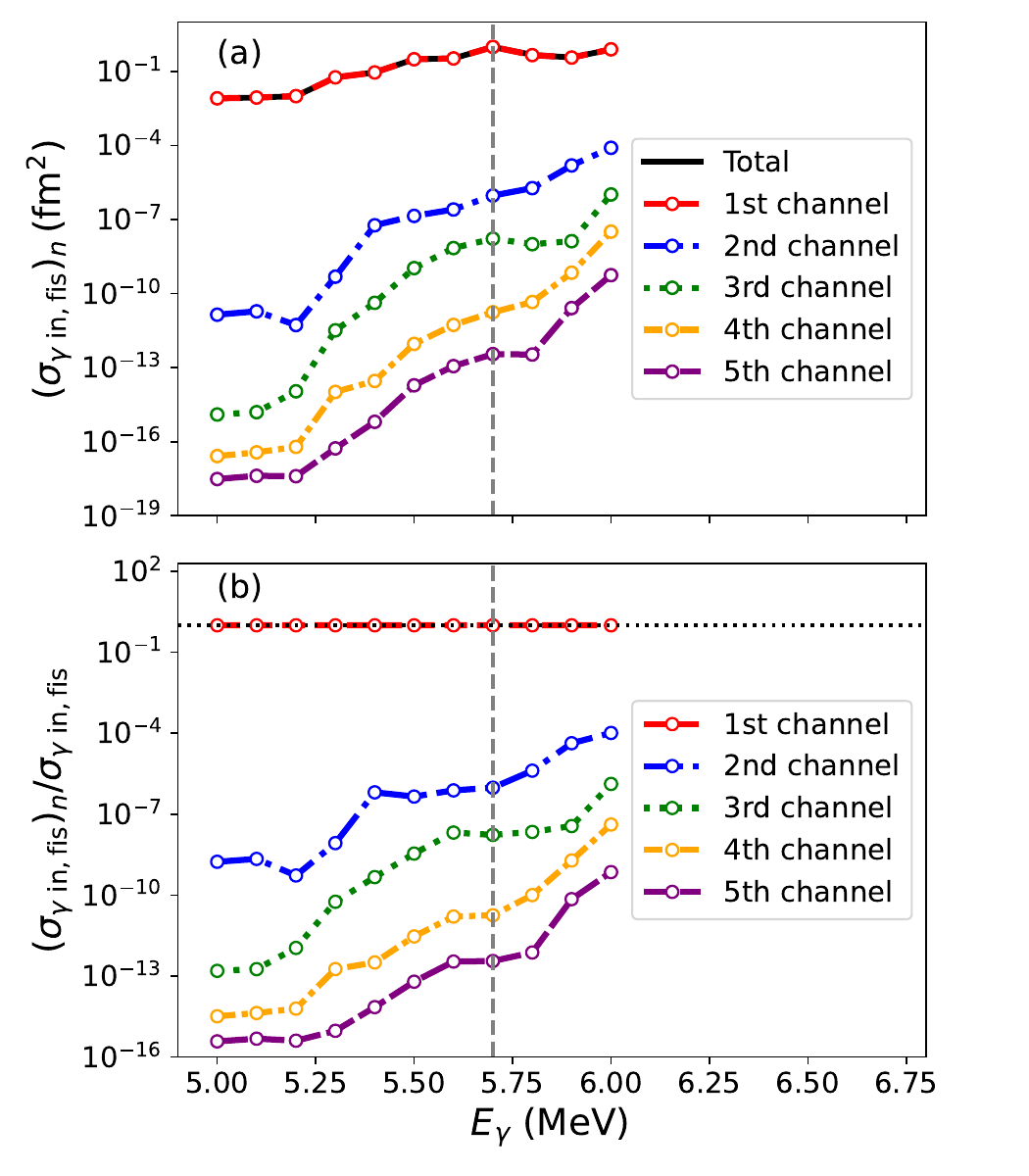}
\caption{(a) Photo-induced fission cross section decomposed into eigenchannels.
The total cross section is the same as that shown in Fig.~\ref{sigma}.
(b) Fractional contribution of each eigenchannel to the total cross section.
The vertical lines indicate the fission barrier height of \(5.7\) MeV, while the horizontal line in panel (b) indicates unity.
}
\label{Eigen}
\end{figure}

In the context of nuclear fission, the eigenchannel decomposition of the Green function in Eq.~(\ref{EIGEN}) has not, to my knowledge, been applied previously. 
Figure~\ref{Eigen}(a) shows the eigenchannel decomposition of the averaged photo-induced fission cross section, obtained by multiplying the kinematic factor by $|t_n|^2$ for the $n$th eigenchannel in Eq.~(\ref{EIGEN}). As can be clearly seen, the first eigenchannel almost completely dominates the fission probability, with $|t_1|^2 \gg |t_n|^2$ for $n \geq 2$.
Figure~\ref{Eigen}(b) shows the ratio of each eigenchannel contribution to the total one.
Although the contributions of the second and higher channels gradually increase with energy, the first channel remains dominant throughout this energy region.

\begin{figure}
\centering
\includegraphics[width=8.6cm]{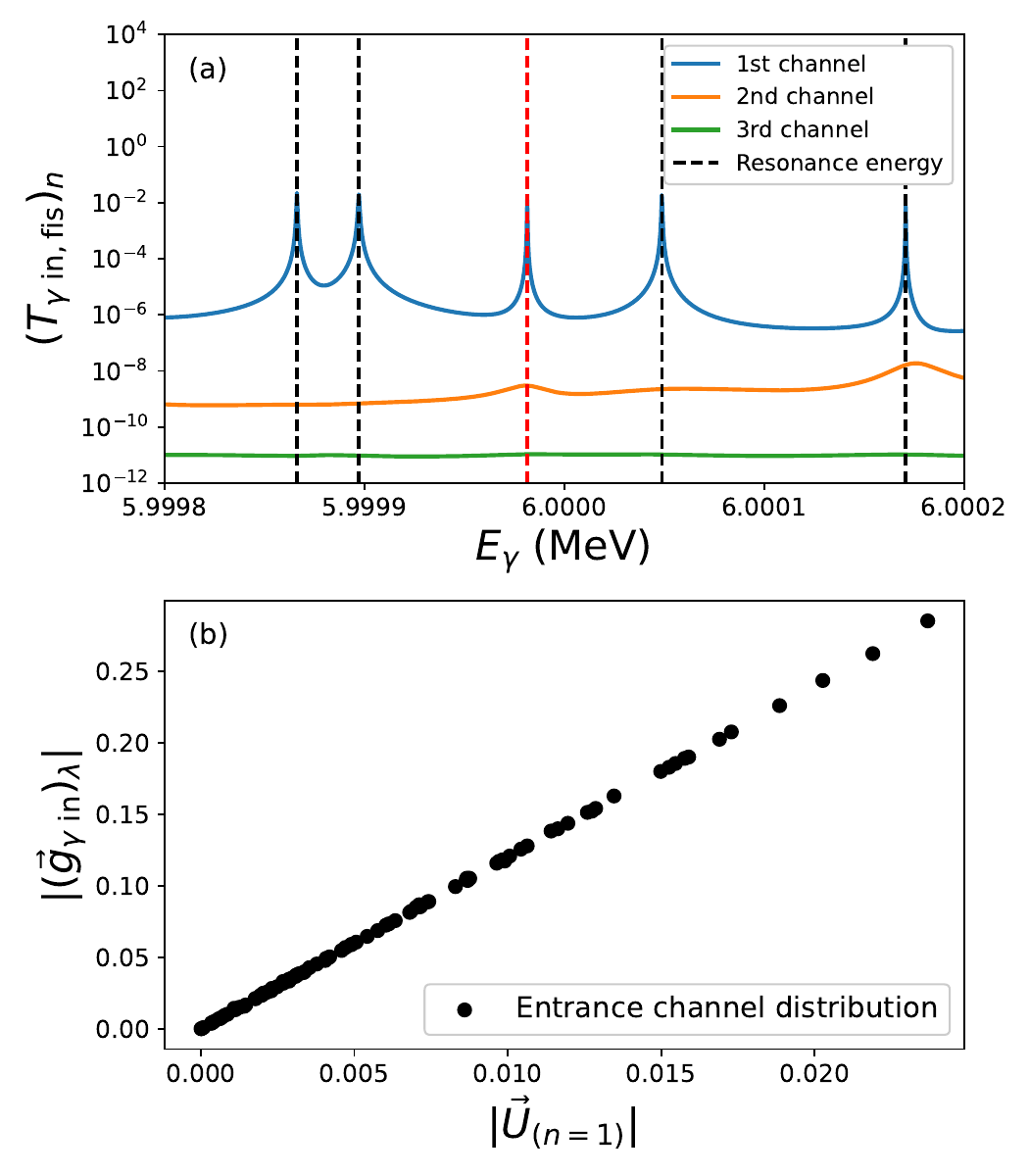}
\caption{
(a) Eigenchannel decomposition of the photo-induced fission transmission coefficient for a fixed random seed. The black vertical lines denote the real parts of the eigenenergies in Eq.~(\ref{HWeq}).
(b) Correlation between the singular vector $U_n(E=6\,\mathrm{MeV})$ for the first eigenchannel ($n=1$) and the $\gamma_\mathrm{in}$-channel component of the GCM collective wave function $(\vec{g}_{\gamma_\mathrm{in}})_\lambda$ for the eigenstate whose eigenenergy $\mathrm{Re}(\tilde{E}_\lambda)$ is closest to $E=6\,\mathrm{MeV}$, as indicated by the red line in panel (a).
}
\label{Eigen_2}
\end{figure}

\begin{figure}
\centering
\includegraphics[width=8.6cm]{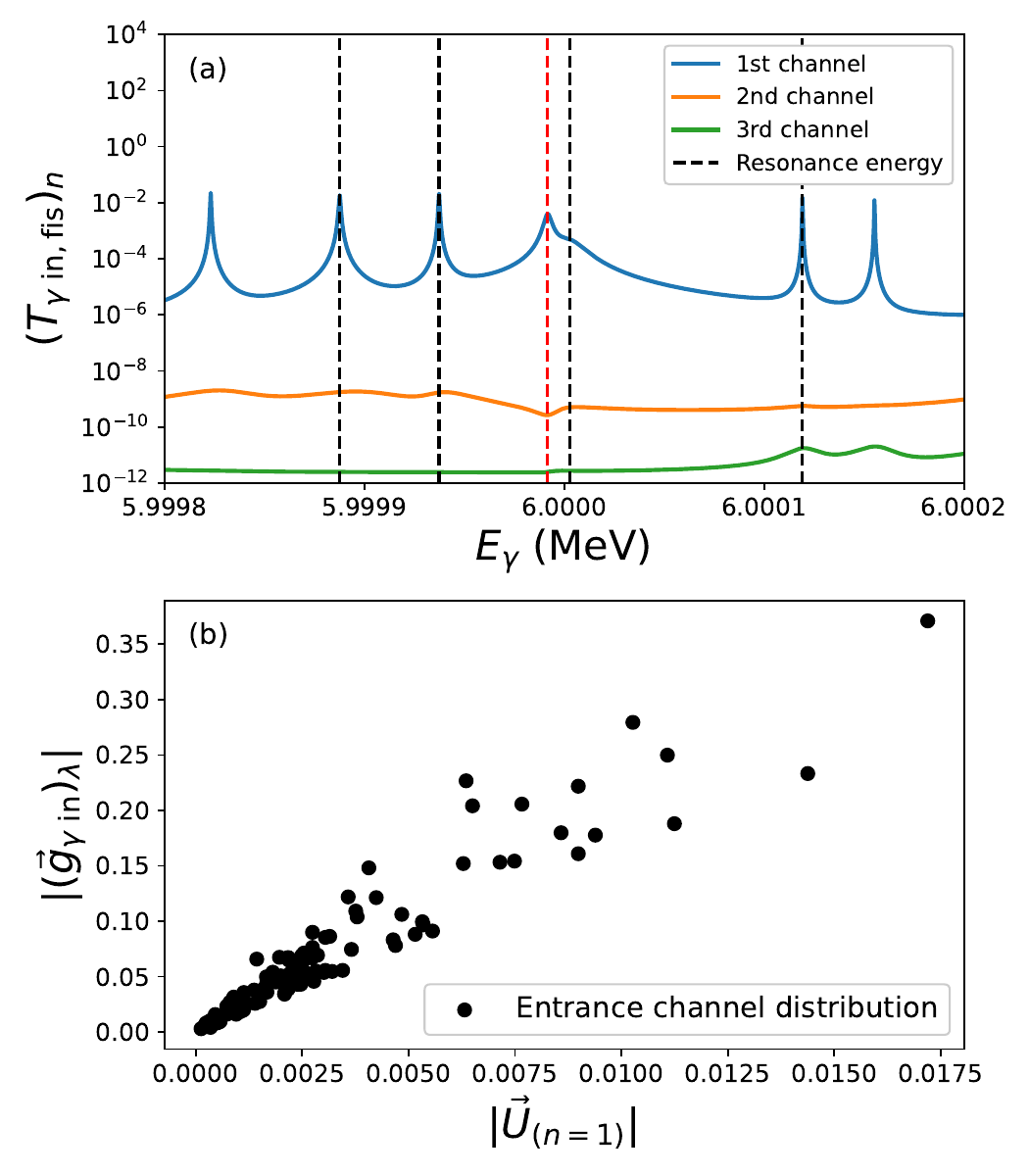}
\caption{
Same as Fig.~\ref{Eigen_2}, but for a different random seed.
}
\label{Eigen_3}
\end{figure}

In addition, I analyze the singular vectors $U$ in Eq.~(\ref{SVD}), which represent how strongly each eigenchannel couples to the individual entrance configurations.
In the present setup, the 100 configurations at $Q_{\min}$ are connected to the photon entrance channel, as in Eq.~(\ref{Gg_in}).
Thus 
$\vec{U}_n$ for the $n$th eigen channel is 100 dimensional vector.

In addition, in the isolated-resonance region, the Green function is well approximated by the contribution from a single resonance eigenstate,
\begin{equation}
    G_{ij}(E)=\sum_\lambda
    \frac{\langle i|\lambda\rangle \langle \tilde{\lambda}|j\rangle}{E-\tilde{E}_\lambda}
    \simeq
    \frac{\langle i|\lambda'\rangle \langle \tilde{\lambda'}|j\rangle}{E-\tilde{E}_{\lambda'}},
    \label{Green_SD}
\end{equation}
where $\lambda'$ denotes the eigenstate whose real part of the eigenenergy is closest to the excitation energy $E$.
Therefore, $\langle i|\lambda'\rangle$ also provides a measure of the coupling strength to the incident channel. 
In the orthogonalized basis after the transformation in Eq.~(\ref{Lowdin}), 
$\langle i|\lambda'\rangle = g_{\lambda'}(i)$ for $i \in \gamma_{\rm in}$ is a 100-dimensional vector in the present setup. 
Mathematically, $\vec{U}_n$ and $g_{\lambda'}(i) \ (i \in \gamma_{\rm in})$ are different quantities, although they have similar physical meanings.

Figure~\ref{Eigen_2}(b) shows the correlation between $\vec{U}_{n=1}(E_\gamma=6\,\mathrm{MeV})$, corresponding to the first eigenchannel, and $|g_{\lambda'}(i)|$ for $i \in \gamma_{\rm in}$, where the eigenenergy of $\lambda'$ is close to $6\,\mathrm{MeV}$. The eigenstate $\lambda'$ corresponds to the peak indicated by the red line in Fig.~\ref{Eigen_2}(a).
Here, neither the energy average nor the ensemble average is performed, 
and the resonance structure remains visible in Fig.~\ref{Eigen_2}(a). 
A clear correlation in Fig.~\ref{Eigen_2}(b) indicates the similarity between the two quantities.

On the other hand, in Fig.~\ref{Eigen_3}(a), 
a different random seed is used, and two resonances overlap around $E=6$ MeV. 
In this case, the correlation between $\vec{U}_{1}(E=6\,\mathrm{MeV})$ and $|g_{\lambda'}(i)|$ for $i \in \gamma_{\rm in}$, shown in Fig.~\ref{Eigen_3}(b), is disturbed. This is because $|g_{\lambda'}(i)|$ is the eigensolution of Eq.~(\ref{HWeq}) and therefore contains information only on a single resonance. In contrast, $\vec{U}_{n}$ incorporates the interference between two or more resonances. Therefore, when resonances are overlaped, the analysis based on $\vec{U}$ is expected to be more appropriate.

The above result, namely that the first eigenchannel dominates the fission probability, is consistent with the transition-state picture \cite{Bohr1939}. In this model, the fission transmission coefficient is given by the sum of the transmission coefficients for the individual transition states,
\begin{equation}
    T_{BW}(E)=\frac{1}{2\pi\rho(E)}\sum_n T_n(E),
\end{equation}
where $n$ labels the transition states. The eigenchannel decomposition similarly provides a decomposition of the photo-induced fission transmission coefficient, as in Eq.~(\ref{EIGEN}), although the dependence on the entrance channel is retained. In the sub-barrier region, fission occurs predominantly through quantum tunneling via the lowest transition state, and the result of the eigenchannel analysis is consistent with this expectation.

The number of degrees of freedom in the fission process is associated with the number of open transition states and is represented by the degree of freedom parameter in the Porter--Thomas distribution of the fission decay widths \cite{Porter1956}.
Recently, the Porter--Thomas distribution in  nutron induced fission was analyzed within the NEGF formalism, and it was shown that the small number of degrees of freedom in the fission process originates from the limited number of Green-function eigenstates that contribute to the fission probability \cite{Uzawa2024}.
Although Ref.~\cite{Uzawa2024} employed a spectral decomposition, whereas an eigenchannel decomposition is used in the present work, both analyses support the conclusion that the small number of degrees of freedom in fission is explained by the eigensolutions of the Green function.

\section{Summary and Future Discussion}

In this article, I have formulated the NEGF method for photo-induced fission in $^{236}$U$(\gamma,f)$ and analyzed the theoretical results, with particular emphasis on the fission cross section.

With an appropriate theoretical setup, the experimental fission cross section is reproduced reasonably well, even in the sub-barrier energy region. I have also applied the eigenchannel analysis and found that the first eigenchannel dominates the fission probability. The reproduction of the experimental fission cross-section is encouraging for the future development of the theoretical fission model along this line.

As future developments of the present model, both a more realistic description of the residual interaction and an extension of the deformation parameter space will be important.
While the many-body Hartree--Fock basis is constructed from the Skyrme EDF in the current framework, simplified forms are employed for the residual interaction in Eqs.~(\ref{pair}, \ref{ran}, \ref{eq:H_db}).
Such a treatment, especially the use of a random interaction, may be insufficient in the low excitation energy region as neutron-induced fission of neutron-rich nuclei, which is of particular importance for the $r$-process~\cite{goriely2015}.
To address this issue, a more realistic treatment of the residual interactions will be required, together with an improved treatment of the overlap matrix beyond the approximation used in Eq.~(\ref{OLP_approx}).

It will also be desirable to move beyond the one-dimensional fission path and include configurations in a two- or higher-dimensional deformation space, especially for a description of the fission fragment mass distributions.

With such developments, the present framework is expected to provide a useful microscopic approach including the fission of neutron-rich nuclei, where experiments are challenging and reliable theoretical predictions are required.

\section{ACKNOWLEDGMENTS}
I thank K. Hagino and O. Iwamoto for useful discussions and a careful reading of the manuscript. 
The numerical calculations were performed through 
the use of SQUID at the
Cybermedia Center, Osaka University,
and Pegasus provided by Multidisciplinary Cooperative Research Program in Center for Computational Sciences, University of Tsukuba.

\appendix
\section{level density formula}\label{app.,LD}
The level density in Eq.~(\ref{T_g_out}) is based on the Gilbert--Cameron formula \cite{Gilbert1965}. In this formula, the lower-energy region is described by the constant-temperature model,
\begin{equation}
    \rho_T(U)=\frac{1}{T}\exp\left(\frac{U+\Delta-E_0}{T}\right),
\end{equation}
where 
$U=E-\Delta$ is the pairing subtracted energy,
$T$ is the nuclear temperature, and $E_0$ is the energy shift parameter. 
In this model, $T$ is assumed to be constant, since in the low-excitation-energy region the increase of $T$ is moderate due to pairing effects \cite{Ericson1960}.

On the other hand, in the higher-energy region, the Fermi-gas model is employed:
\begin{equation}
    \rho_F(U)=
    \frac{\exp\!\left(2\sqrt{a(U)U}\right)}
    {12\sqrt{2}\,\sigma(U)\, a(U)^{1/4}U^{5/4}}.
\end{equation}
Here, $a(U)$ is the level-density parameter, and $\sigma(U)$ is the spin-cutoff parameter. The level-density parameter $a(U)$ is affected by the damping of the shell effect with increasing excitation energy and is parametrized as
\begin{equation}
    a(U)=a^*\left(1+E_{sh}\frac{1-\exp(-\gamma U)}{U}\right),
\end{equation}
where $a^*$, $E_{sh}$, and $\gamma$ denote the asymptotic level-density parameter, the shell-correction energy, and the damping factor, respectively.

The matching energy $U_{\rm match}$ is also introduced as a parameter, and the level density is given by
\begin{equation}
\rho(U)=
\begin{cases}
\rho_T(U), & (U < U_{\rm match}),\\
\rho_F(U), & (U \geq U_{\rm match}).
\end{cases}
\end{equation}

\begin{figure}
\centering
\includegraphics[width=8.6cm]{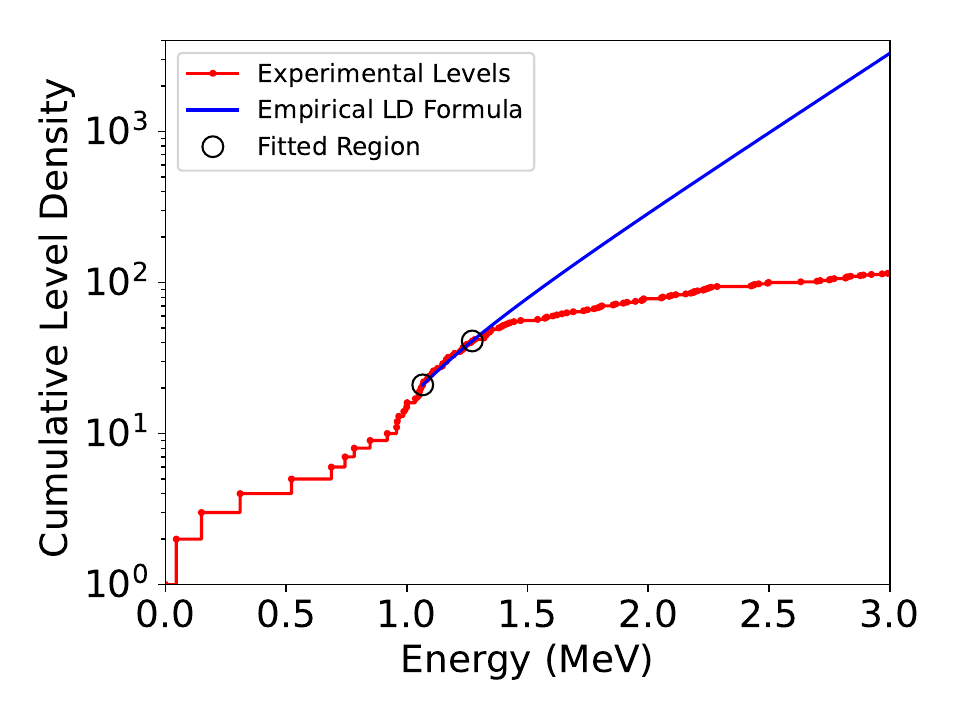}
\caption{
Cumulative number of experimentally known levels in $^{236}$U taken from RIPL-3 \cite{ripl} as a function of excitation energy. 
The solid line shows the cumulative number obtained from the fitted Gilbert--Cameron level-density formula, while the black circles indicate the low-energy levels used in the fit.
}
\label{LD}
\end{figure}

These parameters are determined by fitting to the experimental level-density data. Figure~\ref{LD} shows the cumulative number of levels in $^{236}$U taken from RIPL-3 \cite{ripl}. 
At lower energies, the level sequence is influenced by shell effects, whereas at higher energies many levels are missing.
Actually at $E>1.5$MeV, the slope of the cumulative level density is reduced.
For this reason, it is desirable to 
fit the the level-density parameters whthin the intermediate-energy region indicated by the black circles in the figure. The fitted parameters are summarized in Table~\ref{tab:ldparam}.

For the calculation of the $E1$ transition transmission coefficients, the spin- and parity-dependent level density
\begin{equation}
    \rho(U,J,\pi)=\rho(U)\rho(J)\rho(\pi)
\end{equation}
is required. 
For the parity dependence, we assume \(\rho(\pi)=1/2\), while the spin-dependent part is given by \cite{iwamoto2016}
\begin{equation}
    \rho(J)=\frac{1}{2}\frac{2J+1}{2\sigma^2}\exp\left(-\frac{(J+\frac{1}{2})^2}{2\sigma^2}\right).
\end{equation}
The spin-cutoff parameter is parametrized as 
\begin{equation}
    \sigma=0.01389
A^{5/3}T,
\end{equation}
with the nuclear temperature $T$ defined by
\begin{equation}
    T(U)=\left(\frac{1}{\rho_F(U)}\frac{d\rho_F(U)}{dU}\right)^{-1}.
\end{equation}
This parametrization is adopted from Ref. \cite{MENGONI1994}.

\begin{table}[b]
\caption{The parameters used for the level-density function in Eq.~(\ref{T_g_out}).}
\label{tab:ldparam}
\begin{ruledtabular}
\begin{tabular}{cccccc}
$a^*$ & $\Delta$ & $E_{\mathrm{sh}}$ & $T$ & $E_0$ & $U_{\mathrm{match}}$ \\
(MeV$^{-1}$) & (MeV) & (MeV) & (MeV) & (MeV) & (MeV) \\
\hline
27.42 & 1.56 & 2.75 & 4.15 & -0.36 & 3.98 \\
\end{tabular}
\end{ruledtabular}
\end{table}


\bibliography{references}

@article{Hagino2012,
    author = {Hagino, Kouichi and Takigawa, Noboru},
    title = {Subbarrier Fusion Reactions and Many-Particle Quantum Tunneling},
    journal = {Progress of Theoretical Physics},
    volume = {128},
    number = {6},
    pages = {1061-1106},
    year = {2012},
    month = {12},
    issn = {0033-068X},
    doi = {10.1143/PTP.128.1061},
    url = {https://doi.org/10.1143/PTP.128.1061}
}

@article{MENGONI1994,
author = {Alberto Mengoni and Yutaka Nakajima},
title = {Fermi-Gas Model Parametrization of Nuclear Level Density},
journal = {Journal of Nuclear Science and Technology},
volume = {31},
number = {2},
pages = {151--162},
year = {1994},
publisher = {Taylor \& Francis},
doi = {10.1080/18811248.1994.9735131},


URL = { 
    
        https://doi.org/10.1080/18811248.1994.9735131
    
    

}
}

@article{Paulsson2007,
  title = {Transmission eigenchannels from nonequilibrium Green's functions},
  author = {Paulsson, Magnus and Brandbyge, Mads},
  journal = {Phys. Rev. B},
  volume = {76},
  issue = {11},
  pages = {115117},
  numpages = {7},
  year = {2007},
  month = {Sep},
  publisher = {American Physical Society},
  doi = {10.1103/PhysRevB.76.115117},
  url = {https://link.aps.org/doi/10.1103/PhysRevB.76.115117}
}

@article{Beard2012,
  title = {Photonuclear and radiative-capture reaction rates for nuclear astrophysics and transmutation: $^{92-100}\mathrm{Mo}$, $^{88}\mathrm{Sr}$, $^{90}\mathrm{Zr}$, and $^{139}\mathrm{La}$},
  author = {Beard, M. and Frauendorf, S. and K\"ampfer, B. and Schwengner, R. and Wiescher, M.},
  journal = {Phys. Rev. C},
  volume = {85},
  issue = {6},
  pages = {065808},
  numpages = {19},
  year = {2012},
  month = {Jun},
  publisher = {American Physical Society},
  doi = {10.1103/PhysRevC.85.065808},
  url = {https://link.aps.org/doi/10.1103/PhysRevC.85.065808}
}

@article{Soldatov1995,
  title={Yield and cross section of $^{232}\mathrm{Th}$ and $^{236}\mathrm{U}$ fission induced by $\gamma$ quanta with energies up to 11 MeV},
  author={Soldatov, AS and Smirenkin, GN},
  journal={Physics of Atomic Nuclei},
  volume={58},
  number={2},
  year={1995}
}

@article{Yester1986,
title = {Photofission cross sections of 232Th and 236U from threshold to 8 MeV},
journal = {Nuclear Physics A},
volume = {206},
number = {3},
pages = {593-613},
year = {1973},
issn = {0375-9474},
doi = {https://doi.org/10.1016/0375-9474(73)90089-4},
url = {https://www.sciencedirect.com/science/article/pii/0375947473900894},
author = {M.V. Yester and R.A. Anderm and R.C. Morrison}
}

@article{Caldwell1980,
  title = {Giant resonance for the actinide nuclei: Photoneutron and photofission cross sections for $^{235}\mathrm{U}$, $^{236}\mathrm{U}$, $^{238}\mathrm{U}$, and $^{232}\mathrm{Th}$},
  author = {Caldwell, J. T. and Dowdy, E. J. and Berman, B. L. and Alvarez, R. A. and Meyer, P.},
  journal = {Phys. Rev. C},
  volume = {21},
  issue = {4},
  pages = {1215--1231},
  numpages = {0},
  year = {1980},
  month = {Apr},
  publisher = {American Physical Society},
  doi = {10.1103/PhysRevC.21.1215},
  url = {https://link.aps.org/doi/10.1103/PhysRevC.21.1215}
}

@article{Lynn1980,
  title = {The double-humped fission barrier},
  author = {Bj\o{}rnholm, S. and Lynn, J. E.},
  journal = {Rev. Mod. Phys.},
  volume = {52},
  issue = {4},
  pages = {725--931},
  numpages = {0},
  year = {1980},
  month = {Oct},
  publisher = {American Physical Society},
  doi = {10.1103/RevModPhys.52.725},
  url = {https://link.aps.org/doi/10.1103/RevModPhys.52.725}
}

@Book{VANDENBOSCH1973,
  author    = {Robert Vandenbosch and John R. Huizenga},
  title     = {Nuclear Fission},
  year      = {1973},
  publisher = {Academic Press},
  isbn      = {978-0521599436},
}

@article{Reinhard2021,
title = {The {A}xial {H}artree–{F}ock + {BCS} {C}ode {S}ky{A}x},
journal = {Comput. Phys. Commun.},
volume = {258},
pages = {107603},
year = {2021},
issn = {0010-4655},
doi = {https://doi.org/10.1016/j.cpc.2020.107603},
url = {https://www.sciencedirect.com/science/article/pii/S0010465520302927},
author = {P.-G. Reinhard and B. Schuetrumpf and J.A. Maruhn}
}

@article{Uzawa2024-2,
  title = {Application of the shift-invert {L}anczos algorithm to a nonequilibrium {G}reen's function for transport problems},
  author = {Uzawa, K. and Hagino, K.},
  journal = {Phys. Rev. E},
  volume = {110},
  issue = {5},
  pages = {055302},
  numpages = {8},
  year = {2024},
  month = {Nov},
  publisher = {American Physical Society},
  doi = {10.1103/PhysRevE.110.055302},
  url = {https://link.aps.org/doi/10.1103/PhysRevE.110.055302}
}

@Book{Datta1995,
  author    = {Datta, Supriyo},
  title     = {Electronic Transport in Mesoscopic Systems},
  year      = {1995},
  publisher = {Cambridge University Press, Cambridge},
  isbn      = {978-0521599436},
}

@article{Uzawa2025,
  title = {Microscopic calculation of fission cross sections with the nonequilibrium Green's function method},
  author = {Uzawa, K. and Hagino, K.},
  journal = {Phys. Rev. C},
  volume = {112},
  issue = {1},
  pages = {014326},
  numpages = {12},
  year = {2025},
  month = {Jul},
  publisher = {American Physical Society},
  doi = {10.1103/dyhr-jbds},
  url = {https://link.aps.org/doi/10.1103/dyhr-jbds}
}

@article{Thoss2018,
    author = {Thoss, Michael and Evers, Ferdinand},
    title = {Perspective: Theory of quantum transport in molecular junctions},
    journal = {The Journal of Chemical Physics},
    volume = {148},
    number = {3},
    pages = {030901},
    year = {2018},
    month = {01},
    issn = {0021-9606},
    doi = {10.1063/1.5003306},
    url = {https://doi.org/10.1063/1.5003306},
}

@article{Ozaki2010,
  title = {Efficient implementation of the nonequilibrium Green function method for electronic transport calculations},
  author = {Ozaki, Taisuke and Nishio, Kengo and Kino, Hiori},
  journal = {Phys. Rev. B},
  volume = {81},
  issue = {3},
  pages = {035116},
  numpages = {19},
  year = {2010},
  month = {Jan},
  publisher = {American Physical Society},
  doi = {10.1103/PhysRevB.81.035116},
  url = {https://link.aps.org/doi/10.1103/PhysRevB.81.035116}
}

@book{Datta2005, 
place={Cambridge}, 
title={Quantum Transport: Atom to Transistor}, 
publisher={Cambridge University Press}, 
author={Datta, Supriyo}, 
year={2005},
}

@article{Ericson1960,
author = {Torleif Ericson},
title = {The statistical model and nuclear level densities},
journal = {Advances in Physics},
volume = {9},
number = {36},
pages = {425--511},
year = {1960},
publisher = {Taylor \& Francis},
doi = {10.1080/00018736000101239},


URL = { 
    
        https://doi.org/10.1080/00018736000101239
    
    

},
eprint = { 
    
        https://doi.org/10.1080/00018736000101239
    
    

}

}

@article{goriely2015,
  title={The fundamental role of fission during r-process nucleosynthesis in neutron star mergers},
  author={Goriely, S.},
  journal={The European Physical Journal A},
  volume={51},
  pages={1--21},
  year={2015},
  publisher={Springer}
}

@article{Bender2020,
doi = {10.1088/1361-6471/abab4f},
url = {https://dx.doi.org/10.1088/1361-6471/abab4f},
year = {2020},
month = {oct},
publisher = {IOP Publishing},
volume = {47},
number = {11},
pages = {113002},
author = {M. Bender and others},
title = {Future of nuclear fission theory},
journal = {Journal of Physics G: Nuclear and Particle Physics}
}

@article{Bohr1939,
  title = {The Mechanism of Nuclear Fission},
  author = {Bohr, N. and Wheeler, J. A.},
  journal = {Phys. Rev.},
  volume = {56},
  issue = {5},
  pages = {426--450},
  numpages = {0},
  year = {1939},
  month = {Sep},
  publisher = {American Physical Society},
  doi = {10.1103/PhysRev.56.426},
  url = {https://link.aps.org/doi/10.1103/PhysRev.56.426}
}

@article{Brandbyge2002,
  title = {Density-functional method for nonequilibrium electron transport},
  author = {Brandbyge, Mads and Mozos, Jos\'e-Luis and Ordej\'on, Pablo and Taylor, Jeremy and Stokbro, Kurt},
  journal = {Phys. Rev. B},
  volume = {65},
  issue = {16},
  pages = {165401},
  numpages = {17},
  year = {2002},
  month = {Mar},
  publisher = {American Physical Society},
  doi = {10.1103/PhysRevB.65.165401},
  url = {https://link.aps.org/doi/10.1103/PhysRevB.65.165401}
}

@article{Damle2001,
  title = {Unified description of molecular conduction:  From molecules to metallic wires},
  author = {Damle, P. S. and Ghosh, A. W. and Datta, S.},
  journal = {Phys. Rev. B},
  volume = {64},
  issue = {20},
  pages = {201403},
  numpages = {4},
  year = {2001},
  month = {Oct},
  publisher = {American Physical Society},
  doi = {10.1103/PhysRevB.64.201403},
  url = {https://link.aps.org/doi/10.1103/PhysRevB.64.201403}
}

@article{Bertsch2023,
  title = {Modeling fission dynamics at the barrier in a discrete-basis formalism},
  author = {Bertsch, G. F. and Hagino, K.},
  journal = {Phys. Rev. C},
  volume = {107},
  issue = {4},
  pages = {044615},
  numpages = {11},
  year = {2023},
  month = {Apr},
  publisher = {American Physical Society},
  doi = {10.1103/PhysRevC.107.044615},
  url = {https://link.aps.org/doi/10.1103/PhysRevC.107.044615}
}

@article{Uzawa2024,
  title = {Nonequilibrium {G}reen's function approach to low-energy fission dynamics: Fluctuations in fission reactions},
  author = {Uzawa, K. and Hagino, K.},
  journal = {Phys. Rev. C},
  volume = {110},
  issue = {1},
  pages = {014321},
  numpages = {11},
  year = {2024},
  month = {Jul},
  publisher = {American Physical Society},
  doi = {10.1103/PhysRevC.110.014321},
  url = {https://link.aps.org/doi/10.1103/PhysRevC.110.014321}
}

@Book{ring,
  author    = {P. Ring and P. Schuck},
  title     = {The Nuclear Many-Body Problem},
  year      = {2000},
  publisher = {Springer-Verlag, Berlin},
  isbn      = {978-3-540-21206-5},
}

@article{Leal1999,
author = {Leal,L. C. and H. Derrien and N. M. Larson and R. Q. Wright},
title = {R-{M}atrix {A}nalysis of $^{235}${U} {N}eutron {T}ransmission and {C}ross-{S}ection {M}easurements in the 0- to 2.25-ke{V} {E}nergy {R}ange},
journal = {Nucl. Sci. Eng.},
volume = {131},
number = {2},
pages = {230-253},
year = {1999},
publisher = {Taylor & Francis},
doi = {10.13182/NSE99-A2031},
URL = { https://doi.org/10.13182/NSE99-A2031}
}

@article{Barranco1990,
title = {Large-amplitude motion in superfluid fermi droplets},
journal = {Nuclear Physics A},
volume = {512},
number = {2},
pages = {253-274},
year = {1990},
issn = {0375-9474},
doi = {https://doi.org/10.1016/0375-9474(90)93232-U},
url = {https://www.sciencedirect.com/science/article/pii/037594749093232U},
author = {F. Barranco and G. F. Bertsch and R. A. Broglia and E. Vigezzi}
}

@article{Bush1992,
  title = {Shape diffusion in the shell model},
  author = {Bush, B. W. and Bertsch, G. F. and Brown, B. A.},
  journal = {Phys. Rev. C},
  volume = {45},
  issue = {4},
  pages = {1709--1719},
  numpages = {0},
  year = {1992},
  month = {Apr},
  publisher = {American Physical Society},
  doi = {10.1103/PhysRevC.45.1709},
  url = {https://link.aps.org/doi/10.1103/PhysRevC.45.1709}
}

@article{Uzawa2023,
  title = {Schematic model for induced fission in a configuration-interaction approach},
  author = {Uzawa, K. and Hagino, K.},
  journal = {Phys. Rev. C},
  volume = {108},
  issue = {2},
  pages = {024319},
  numpages = {9},
  year = {2023},
  publisher = {American Physical Society},
  doi = {10.1103/PhysRevC.108.024319},
  url = {https://link.aps.org/doi/10.1103/PhysRevC.108.024319}
}

@article{Feshbach1958,
title = {Unified theory of nuclear reactions},
journal = {Annals of Physics},
volume = {5},
number = {4},
pages = {357-390},
year = {1958},
issn = {0003-4916},
doi = {https://doi.org/10.1016/0003-4916(58)90007-1},
url = {https://www.sciencedirect.com/science/article/pii/0003491658900071},
author = {Herman Feshbach}
}

@article{Gilbert1965,
  title={A composite nuclear-level density formula with shell corrections},
  author={Gilbert, Arnold and Cameron, Alastair Graham Walter},
  journal={Canadian Journal of Physics},
  volume={43},
  number={8},
  pages={1446--1496},
  year={1965},
  publisher={NRC Research Press Ottawa, Canada}
}

@article{ripl,
title = {{RIPL} – Reference Input Parameter Library for Calculation of Nuclear Reactions and Nuclear Data Evaluations},
journal = {Nuclear Data Sheets},
volume = {110},
number = {12},
pages = {3107-3214},
year = {2009},
note = {Special Issue on Nuclear Reaction Data},
issn = {0090-3752},
doi = {https://doi.org/10.1016/j.nds.2009.10.004},
url = {https://www.sciencedirect.com/science/article/pii/S0090375209000994},
author = {R. Capote and M. Herman and P. Obložinský and P.G. Young and S. Goriely and T. Belgya and A.V. Ignatyuk and A.J. Koning and S. Hilaire and V.A. Plujko and M. Avrigeanu and O. Bersillon and M.B. Chadwick and T. Fukahori and Zhigang Ge and Yinlu Han and S. Kailas and J. Kopecky and V.M. Maslov and G. Reffo and M. Sin and E.Sh. Soukhovitskii and P. Talou}
}

@article{Alhassid2021,
title = {Addendum to “Derivation of {K}-matrix reaction theory in a discrete basis formalism” [Ann. Phys. 419 (2020) 168233]},
journal = {Annals of Physics},
volume = {424},
pages = {168381},
year = {2021},
issn = {0003-4916},
doi = {https://doi.org/10.1016/j.aop.2020.168381},
url = {https://www.sciencedirect.com/science/article/pii/S0003491620303158},
author = {Y. Alhassid and G. F. Bertsch and P. Fanto}
}

@article{Bulgac2016,
  title = {Induced Fission of $^{240}\mathrm{Pu}$ within a Real-Time Microscopic Framework},
  author = {Bulgac, Aurel and Magierski, Piotr and Roche, Kenneth J. and Stetcu, Ionel},
  journal = {Phys. Rev. Lett.},
  volume = {116},
  issue = {12},
  pages = {122504},
  numpages = {7},
  year = {2016},
  month = {Mar},
  publisher = {American Physical Society},
  doi = {10.1103/PhysRevLett.116.122504},
  url = {https://link.aps.org/doi/10.1103/PhysRevLett.116.122504}
}

@article{Hagino2022,
  title = {Diabatic Hamiltonian matrix elements made simple},
  author = {Hagino, K. and Bertsch, G. F.},
  journal = {Phys. Rev. C},
  volume = {105},
  issue = {3},
  pages = {034323},
  numpages = {6},
  year = {2022},
  month = {Mar},
  publisher = {American Physical Society},
  doi = {10.1103/PhysRevC.105.034323},
  url = {https://link.aps.org/doi/10.1103/PhysRevC.105.034323}
}

@article{iwamoto2016,
title = {The CCONE Code System and its Application to Nuclear Data Evaluation for Fission and Other Reactions},
journal = {Nuclear Data Sheets},
volume = {131},
pages = {259-288},
year = {2016},
note = {Special Issue on Nuclear Reaction Data},
issn = {0090-3752},
doi = {https://doi.org/10.1016/j.nds.2015.12.004},
url = {https://www.sciencedirect.com/science/article/pii/S0090375215000691},
author = {O. Iwamoto and N. Iwamoto and S. Kunieda and F. Minato and K. Shibata}
}

@article{Kortelainen2012,
  title = {Nuclear energy density optimization: Large deformations},
  author = {Kortelainen, M. and McDonnell, J. and Nazarewicz, W. and Reinhard, P.-G. and Sarich, J. and Schunck, N. and Stoitsov, M. V. and Wild, S. M.},
  journal = {Phys. Rev. C},
  volume = {85},
  issue = {2},
  pages = {024304},
  numpages = {15},
  year = {2012},
  month = {Feb},
  publisher = {American Physical Society},
  doi = {10.1103/PhysRevC.85.024304},
  url = {https://link.aps.org/doi/10.1103/PhysRevC.85.024304}
}

@article{Porter1956,
  title = {Fluctuations of Nuclear Reaction Widths},
  author = {Porter, C. E. and Thomas, R. G.},
  journal = {Phys. Rev.},
  volume = {104},
  issue = {2},
  pages = {483--491},
  numpages = {0},
  year = {1956},
  month = {Oct},
  publisher = {American Physical Society},
  doi = {10.1103/PhysRev.104.483},
  url = {https://link.aps.org/doi/10.1103/PhysRev.104.483}
}

@article{Negele1978,
  title = {Dynamics of induced fission},
  author = {Negele, J. W. and Koonin, S. E. and M\"oller, P. and Nix, J. R. and Sierk, A. J.},
  journal = {Phys. Rev. C},
  volume = {17},
  issue = {3},
  pages = {1098--1115},
  numpages = {0},
  year = {1978},
  month = {Mar},
  publisher = {American Physical Society},
  doi = {10.1103/PhysRevC.17.1098},
  url = {https://link.aps.org/doi/10.1103/PhysRevC.17.1098}
}

@article{Simenel2014,
  title = {Formation and dynamics of fission fragments},
  author = {Simenel, C. and Umar, A. S.},
  journal = {Phys. Rev. C},
  volume = {89},
  issue = {3},
  pages = {031601},
  numpages = {5},
  year = {2014},
  month = {Mar},
  publisher = {American Physical Society},
  doi = {10.1103/PhysRevC.89.031601},
  url = {https://link.aps.org/doi/10.1103/PhysRevC.89.031601}
}

@article{Goddard2015,
  title = {Fission dynamics within time-dependent Hartree-Fock: Deformation-induced fission},
  author = {Goddard, Philip and Stevenson, Paul and Rios, Arnau},
  journal = {Phys. Rev. C},
  volume = {92},
  issue = {5},
  pages = {054610},
  numpages = {20},
  year = {2015},
  month = {Nov},
  publisher = {American Physical Society},
  doi = {10.1103/PhysRevC.92.054610},
  url = {https://link.aps.org/doi/10.1103/PhysRevC.92.054610}
}

@article{Scamps2018,
  title={Impact of pear-shaped fission fragments on mass-asymmetric fission in actinides},
  author={Scamps, Guillaume and Simenel, C{\'e}dric},
  journal={Nature},
  volume={564},
  number={7736},
  pages={382--385},
  year={2018},
  publisher={Nature Publishing Group UK London}
}

@article{Bertsch2020,
  title = {Schematic reaction-theory model for nuclear fission},
  author = {Bertsch, G. F.},
  journal = {Phys. Rev. C},
  volume = {101},
  issue = {3},
  pages = {034617},
  numpages = {5},
  year = {2020},
  month = {Mar},
  publisher = {American Physical Society},
  doi = {10.1103/PhysRevC.101.034617},
  url = {https://link.aps.org/doi/10.1103/PhysRevC.101.034617}
}

@article{Schunck2023,
  title = {Microscopic calculation of fission product yields for odd-mass nuclei},
  author = {Schunck, N. and Verriere, M. and Potel Aguilar, G. and Malone, R. C. and Silano, J. A. and Ramirez, A. P. D. and Tonchev, A. P.},
  journal = {Phys. Rev. C},
  volume = {107},
  issue = {4},
  pages = {044312},
  numpages = {17},
  year = {2023},
  month = {Apr},
  publisher = {American Physical Society},
  doi = {10.1103/PhysRevC.107.044312},
  url = {https://link.aps.org/doi/10.1103/PhysRevC.107.044312}
}

@article{Zhao2022,
  title = {Time-dependent generator coordinate method study of fission. II. Total kinetic energy distribution},
  author = {Zhao, Jie and Nik\ifmmode \check{s}\else \v{s}\fi{}i\ifmmode \acute{c}\else \'{c}\fi{}, Tamara and Vretenar, Dario},
  journal = {Phys. Rev. C},
  volume = {106},
  issue = {5},
  pages = {054609},
  numpages = {6},
  year = {2022},
  month = {Nov},
  publisher = {American Physical Society},
  doi = {10.1103/PhysRevC.106.054609},
  url = {https://link.aps.org/doi/10.1103/PhysRevC.106.054609}
}

@article{Zhao2021,
  title = {Time-dependent generator coordinate method study of fission. II. Total kinetic energy distribution},
  author = {Zhao, Jie and Nik\ifmmode \check{s}\else \v{s}\fi{}i\ifmmode \acute{c}\else \'{c}\fi{}, Tamara and Vretenar, Dario},
  journal = {Phys. Rev. C},
  volume = {106},
  issue = {5},
  pages = {054609},
  numpages = {6},
  year = {2022},
  month = {Nov},
  publisher = {American Physical Society},
  doi = {10.1103/PhysRevC.106.054609},
  url = {https://link.aps.org/doi/10.1103/PhysRevC.106.054609}
}

@ARTICLE{Verriere2020,
    
AUTHOR={Verriere, Marc  and Regnier, David },
           
TITLE={The Time-Dependent Generator Coordinate Method in Nuclear Physics},
          
JOURNAL={Frontiers in Physics},
          
VOLUME={Volume 8 - 2020},
  
YEAR={2020},
  
URL={https://www.frontiersin.org/journals/physics/articles/10.3389/fphy.2020.00233},
  
DOI={10.3389/fphy.2020.00233},
  
ISSN={2296-424X}}

@article{Tao2017,
  title = {Microscopic study of induced fission dynamics of $^{226}\mathrm{Th}$ with covariant energy density functionals},
  author = {Tao, H. and Zhao, J. and Li, Z. P. and Nik\ifmmode \check{s}\else \v{s}\fi{}i\ifmmode \acute{c}\else \'{c}\fi{}, T. and Vretenar, D.},
  journal = {Phys. Rev. C},
  volume = {96},
  issue = {2},
  pages = {024319},
  numpages = {10},
  year = {2017},
  month = {Aug},
  publisher = {American Physical Society},
  doi = {10.1103/PhysRevC.96.024319},
  url = {https://link.aps.org/doi/10.1103/PhysRevC.96.024319}
}

@article{Zdeb2017,
  title = {Fission dynamics of $^{252}\mathbf{Cf}$},
  author = {Zdeb, A. and Dobrowolski, A. and Warda, M.},
  journal = {Phys. Rev. C},
  volume = {95},
  issue = {5},
  pages = {054608},
  numpages = {7},
  year = {2017},
  month = {May},
  publisher = {American Physical Society},
  doi = {10.1103/PhysRevC.95.054608},
  url = {https://link.aps.org/doi/10.1103/PhysRevC.95.054608}
}

@article{Regnier2016,
  title = {Fission fragment charge and mass distributions in $^{239}\mathrm{Pu}(n,f)$ in the adiabatic nuclear energy density functional theory},
  author = {Regnier, D. and Dubray, N. and Schunck, N. and Verri\`ere, M.},
  journal = {Phys. Rev. C},
  volume = {93},
  issue = {5},
  pages = {054611},
  numpages = {15},
  year = {2016},
  month = {May},
  publisher = {American Physical Society},
  doi = {10.1103/PhysRevC.93.054611},
  url = {https://link.aps.org/doi/10.1103/PhysRevC.93.054611}
}

@article{Axel1962,
  title = {Electric Dipole Ground-State Transition Width Strength Function and 7-Mev Photon Interactions},
  author = {Axel, Peter},
  journal = {Phys. Rev.},
  volume = {126},
  issue = {2},
  pages = {671--683},
  numpages = {0},
  year = {1962},
  month = {Apr},
  publisher = {American Physical Society},
  doi = {10.1103/PhysRev.126.671},
  url = {https://link.aps.org/doi/10.1103/PhysRev.126.671}
}

@article{Bernard2011,
  title = {Microscopic and nonadiabatic Schr\"odinger equation derived from the generator coordinate method based on zero- and two-quasiparticle states},
  author = {Bernard, R. and Goutte, H. and Gogny, D. and Younes, W.},
  journal = {Phys. Rev. C},
  volume = {84},
  issue = {4},
  pages = {044308},
  numpages = {20},
  year = {2011},
  month = {Oct},
  publisher = {American Physical Society},
  doi = {10.1103/PhysRevC.84.044308},
  url = {https://link.aps.org/doi/10.1103/PhysRevC.84.044308}
}

@article{Berger1991,
title = {Time-dependent quantum collective dynamics applied to nuclear fission},
journal = {Computer Physics Communications},
volume = {63},
number = {1},
pages = {365-374},
year = {1991},
issn = {0010-4655},
doi = {https://doi.org/10.1016/0010-4655(91)90263-K},
url = {https://www.sciencedirect.com/science/article/pii/001046559190263K},
author = {J.F. Berger and M. Girod and D. Gogny}
}

@article{Goutte2005,
  title = {Microscopic approach of fission dynamics applied to fragment kinetic energy and mass distributions in $^{238}\mathrm{U}$},
  author = {Goutte, H. and Berger, J. F. and Casoli, P. and Gogny, D.},
  journal = {Phys. Rev. C},
  volume = {71},
  issue = {2},
  pages = {024316},
  numpages = {13},
  year = {2005},
  month = {Feb},
  publisher = {American Physical Society},
  doi = {10.1103/PhysRevC.71.024316},
  url = {https://link.aps.org/doi/10.1103/PhysRevC.71.024316}
}

@article{Goddard2016,
  title = {Fission dynamics within time-dependent Hartree-Fock. II. Boost-induced fission},
  author = {Goddard, Philip and Stevenson, Paul and Rios, Arnau},
  journal = {Phys. Rev. C},
  volume = {93},
  issue = {1},
  pages = {014620},
  numpages = {21},
  year = {2016},
  month = {Jan},
  publisher = {American Physical Society},
  doi = {10.1103/PhysRevC.93.014620},
  url = {https://link.aps.org/doi/10.1103/PhysRevC.93.014620}
}

\end{document}